\documentclass[pre,twocolumn,footnote,amsmath,amssymb,floatfix,superscriptaddress]{revtex4}
\usepackage{bm}

\usepackage{amssymb,amsmath}
\usepackage{graphicx}
\usepackage{epstopdf}
\usepackage{epsfig}
\usepackage{color}
\usepackage{amsfonts}
\usepackage{bm}
\usepackage{amscd}
\usepackage{epsfig}
\usepackage{subfigure}
\usepackage{color}

\def\ome{{\mbox{\boldmath{$\omega$}}}}

\newcommand{\be}{\begin{equation}}
\newcommand{\ee}{\end{equation}}
\newcommand{\ave}[1]{\langle #1 \rangle}
\newcommand{\nn}{\nonumber}
\newcommand{\ov}[1]{\overline{#1}}

\begin{document}
\title{Linear and Nonlinear Mesoscopic Thermoelectric Transport with Coupling to Heat Baths}
(Pat. pending, to appear in Comptes Rendus)

\author{Jian-Hua Jiang}\email{joejhjiang@sina.com}
\affiliation{School of Physical Science and Technology, Soochow University, 1
  Shizi Street, Suzhou, Jiangsu, China}
\author{ Yoseph  Imry}\email{yoseph.imry@weizmann.ac.il}
\affiliation{Department of Condensed Matter Physics, Weizmann Institute of Science, Rehovot 76100, Israel}.

\date{\today}


\begin{abstract}
Decades of research on thermoelectrics stimulated by the fact that nano- and
meso-scale thermoelectric transport could yield higher energy conversion efficiency and
output power has recently uncovered a new direction on inelastic thermoelectric
effects. We introduce the history, motivation, and perspectives on mesoscopic inelastic
thermoelectric effects.
\end{abstract}

\maketitle

\section{Introduction}
Research on the science and technology of thermoelectric
phenomenon has a long history since its discovery almost two
centuries ago\cite{honig}. The modern theory for thermoelectric
transport in bulk semiconductors was established, with the help of
energy band theory and semi-classical transport theory, in the middle
of the last century\cite{honig}. The key concepts such as the figure of
merit and the power factor were introduced then, which facilitated and
stimulated many theoretical and 
experimental studies. It was found that the figure of merit (a measure
of the optimal thermoelectric efficiency in a material),
\be
ZT = \frac{\sigma S^2 T}{\kappa}
\ee
is limited by the following interrelated transport quantities: the electrical conductivity
$\sigma$, the Seebeck coefficient $S$, and the thermal conductivity $\kappa$. The latter
has contributions from the electronic transport $\kappa_e$ and other mechanisms (mainly from
phonons in semiconducting materials, $\kappa_p$), i.e., $\kappa=
\kappa_e+\kappa_p$. Increasing $S$ usually leads to reduced electrical conductivity
$\sigma$. In addition, $\sigma$ and $\kappa_e$ are interrelated. In metals,
these two quantities mostly follow the Wiedemann-Franz law
\be
\kappa = LT \sigma,
\ee
where $L\equiv a_L(k_B/e)^2$ is the Lorenz number ($a_L\sim 1$, depending on
the material and the temperature). In semiconductors, similar
relations are usually approximately confirmed. When the Wiedeman-Franz
law holds, the thermoelectric figure of merit is approximately $S^2
e^2 / k_B^2$, which is often smaller than 1.

Mahan and Sofo proposed to improve the figure of merit by using
conductors with very narrow energy bands, so that the thermal
conductivity $\kappa_e$ is reduced from the Wiedemann-Franz
law\cite{ms}. However, this was found lately to be less effective as, for
example, the output power is suppressed in the zero band width
limit\cite{low-diss} due to, {\it e,g.}, suppressed electrical
conductivity\cite{zhou}.

Alongside with those developments, the scientific community was pursuing a better
understanding of (charge) transport in nano- and meso-scale systems,
localization phenomena, and strongly interacting electron systems. These
studies activated research on thermoelectric transport in non-standard
(including bulk) semiconductors, which is still
ongoing\cite{boundary}. The most influential studies 
are based on two seminal works by Hicks and Dresselhaus\cite{hicks}, who found that
in reduced-dimension semiconductors, such as quantum wells (2D), quantum wires
(1D) and quantum dots (0D) heterostructures, the interrelation between
the electrical conductivity, the Seebeck coefficient, and the thermal
conductivity can be partially decoupled. This is mainly because the density of states can be effectively
enhanced and modulated (at different energies) when the spatial confinement along any
direction is small enough to induce significant quantum-confinement effects.
Opportunities for enhancing the thermoelectric figure of merit $ZT$ and the power
factor $S^2\sigma$ by engineering nanostructures and nanomaterials
thus emerge\cite{nano1,nano2,kanatzidis,semimetal,ms-review}. 
In addition, when low dimensional structures are packed up
(densely) to macroscopic structures, the abundant interfaces effectively
reduce the phonon heat conductivity\cite{nano1,nano2}. If the phonon heat conductvity is reduced
more significantly (e.g., by introducing effective phonon scattering centers, or
packing materials with mismatched phonon impedance) then the electrical
conductivity, the thermoelectric figure of merit can be
improved\cite{kanatzidis}.  This was demonstrated in BiTe quantum well
superlattices and PbTe 
quantum dot superlattices\cite{nano2}. In the latter a high density of quantum dots forming
a regular array induces a large electronic density of states in the PbSeTe alloy
layer. Other methods, such as energy filtering and semimetal-semiconductor
transition, in engineering the electronic density of states and the energy
dependent conductivity are also introduced\cite{semimetal,ms-review}. The underlying physics is manifested
in the Mott-Cutler formula\cite{cutler,joe}
\be
S = \frac{\ave{E-\mu}}{eT}, \quad \kappa = \sigma \frac{{\rm
    Var}(E-\mu)}{e^2 T} \label{s-kappa}
\ee
where ${\rm Var}(E-\mu)=\ave{(E-\mu)^2}-\ave{(E-\mu)}^2$ is the
variance of the transport electronic energy; and the averages of powers of
the electronic energy are defined as
\begin{subequations}
\begin{align}
\ave{(E-\mu)^n} & = \frac{1}{\sigma} \int dE \sigma(E) (E-\mu)^n
[-\partial_E f_F(E)],\nn \\ 
\quad &\quad\quad\quad n=1, 2, \\
\sigma & = \int dE \sigma(E) [-\partial_E f_F(E) ] .
\end{align}
\end{subequations}
Here $\sigma(E)$ is the energy dependent conductivity and
$f_F(E)=1/[\exp(\frac{E-\mu}{k_BT})+1]$ is the Fermi distribution
function. The energy dependence of the conductivity $\sigma(E)$ is the
key quantity for engineering the thermoelectric transport
properties. It can be written as $\sigma(E) = ek_BT\rho(E)b(E)$ where
$\rho(E)$ is the density of states (DOS)  and $b(E)$ is the
energy-dependent mobility. 

Nanostructures and nanocomposites have been demonstrated as  effective
ways to tune the carrier density of states and the energy dependence of carrier
scattering, as well as to reduce the phonon thermal
conductivity. These methods lead to high thermoelectric efficiency and
power density for applications. Nevertheless, several nontrivial
aspects of mesoscopic electron transport have not been exploited,
which might give a chance to further enhance thermoelectric efficiency
and power. These are: (1) nonlinear effects, (2) inelastic effects
(due to strong carrier-carrier interaction and carrier-phonon
interaction), (3) thermoelectric transport with spatially separated
electrical and thermal currents. The main purpose of this review
is to emphasize those aspects and their potential values for improving
thermoelectric performance, as well as realization of thermoelectric
diodes and transistors. We mainly focus on the inelastic effects and
show that they can offer an alternative (new) route toward high
performance thermoelectric structures that may reduce the necessity
for novel functional materials. We further discuss nonlinear effects
and the spatial separation of heat and electrical currents in
thermoelectric transport through inelastic thermoelectric
transport. We will illustrate how these aspects may lead to better
thermoelectrics. Several example systems are used to demonstrate the
principles.

In the next section we consider bounds (which, when reached,  are sometimes referred
to as ``quantized values'') of thermoelectric response
coefficients. These bounds determine sometimes \cite{RW} the
scales over which the maximal efficiency is reduced from Carnot. In section III, we
consider linear and nonlinear transport above a barrier. 
The latter is crossed by thermal activation, the energy for which is
taken from the (usually phonons, or the other electrons) associated
heat baths. In section IV we discuss a simple, quite generic, model
\cite{3t,3tjap} for inelastic transport using a heat
bath. Synergistic heat and electrical rectification and
transistor effects in systems with pronounced inelastic
thermoelectric transport\cite{3t-nonl} is
reviewed in section V. We conclude and summarize in section VI.

\section{Bounds and quantized values of thermal-electric transport
  coefficients (This section is dedicated to Jacob
  Bekenstein, a highly esteemed colleague, some of whose early  work is
  reviewed here)}

In this section, we discuss bounds (upper limits) on electric, thermal
and thermoelectric  transport coefficients. 
For the first two, these bounds can be reached in ideal conducting
channels and give rise to ``quantized'' values 
of these coefficients.

It makes physical sense that maximal values of currents carried in a
pipe are proportional to the cross sectional area of that pipe. The
same is valid for the transport of charge, energy and heat through a
conducting wire. This is of interest by itself and because, as found
by Whitney \cite{RW} these bounds also give the scales on which
the maximal efficiency is reduced from Carnot.

We start with the simplest case of charge transport, which is not the
first example to have been  discussed \cite{JB}, but the easiest to
understand \cite{YI}. Consider a 1D wire carrying electrical current,
at a temperature $T \ll E_F$ with $E_F$ being the Fermi energy.
We imagine, following Landauer\cite{Land}, that the strictly 1D wire
(so narrow that the lowest transverse excitation energy is larger than
$E_F$ plus several $k_BT$, so only one transverse state is relevant) 
connecting two reservoirs each at equilibrium, but with slightly
different  electrochemical potentials $\mu_L$ and $\mu_R$ and, at
this stage, identical temperatures, $T$. The current flowing between
L and R is: 
\be
I =  e\int dE (f_L - f_R) v(E) \rho(E), 
\ee
$v(E)$ and $\rho(E)$ being the velocity and the density of states
at energy $E$. $f_L $ and $ f_R$ are the Fermi functions of $L$ and $R$.
Since  the 1D single-particle DOS, spin included, is given by 
\be
\rho(E) = \frac{1}{\pi \hbar v(E)},
\ee
the velocity factors cancel. For {\em linear response} $f_L - f_R = - eV
f^{'}$ with $f^{'}\equiv \partial_E f$ ($f$ is the equilibrium Fermi
distribution), the integral trivially gives $I = \frac{e^2}{\pi \hbar}
V$. Therefore, the conductance is given by
\be 
G = I/V =  \frac{e^2}{\pi \hbar}
\ee
This inconspicuous result is amazingly deep! The conductance of an {\em
  ideal} 1D (``single channel'') wire is universal and totally
independent on material and properties of the wire! No single-channel
wire can have a larger conductance than an ideal one. So, this value
is both an upper limit and a so called ``quantized value'' of this
conductance. The upper limit is realized in the ideal conductor (no
scattering) limit. We shall soon see that a similar phenomenon occurs
for the thermal conductance, but before that we discuss the effect of
a finite thickness wire, comprising several channels. 

If the wire has $n$ transverse states below the Fermi energy, there
will be a continuum of longitudinal (parallel to the wire) states for
each, which will create a $\frac{e^2}{\pi \hbar} $ conductance. Thus,
the total low-temperature conductance will be $n\frac{e^2}{\pi\hbar}$,
Similarly, if the temperature is higher than the
separation, $\Delta$,  to a higher transverse state, its channel will
be populated and it too will contribute another conductance
quantum. Since $\Delta \sim \hbar^2/ma^2$ where $a$ is the relevant
dimension of the wire, the condition $T\gg\Delta $ (still, with  $T\ll
E_F$) is equivalent to $a \gg \lambda_T$. (Similarly, $\Delta \gg E_F$
can also by described by $a \ll k_F^{-1}$). Thus, the number of
channels, $n$, in a wire is on the order of the number of elementary
$k_F^{-2}$ or $\lambda_T^2$ areas at zero or higher temperatures,
respectively. At the higher temperatures, the measured conductance,
related to the number of channels, is a quantum property, dependent on
$\hbar$! This has been mentioned by Whitney \cite{RW} in the context
of thermal conductivity, which is the subject we shall discuss
shortly.

The quantized conductance has been observed experimentally,
concurrently by two groups \cite{exp}, in suitable ballistic quantum
point contacts in GaAs, within a percent accuracy, about two years
after the prediction. More recently, it became a standard item in
mesoscopic Physics. 

Now we discuss the electronic thermal conductance, to that end we
assume that the two Landauer reservoirs have the same electrochemical
potentials and different temperatures $T_L$ and $T_R$. (in linear
response, the effects of $T_L - T_R$ and $\mu_L - \mu_R$ are simply
additive). We want \cite{rem1} the heat current between $L$ and
$R$. Thermodynamics tells that an electron of energy transferring from
$L$ to $R$ carries heat given by \cite{Ziman,Amir} $E-\mu = TS$. Thus,
the heat current, $ \int dE (f_L - f_R) (E-\mu)v(E) \rho(E)$ between
$L$ and $R$, is given by 
\be
J_Q = (T_L - T_R) \int dE (-f^{'}) \frac{(E-\mu)^2}{T} v(E) \rho(E),
\ee
where we used $f_L - f_R = (-f^{'}) \frac{(E-\mu)}{T} (T_L - T_R)$. 
We now use the Sommerfeld expansion (see, e.g. \cite{Ziman})  
$\int_0^\infty dE (E - \mu)^2 (-f^{'}) = (\pi^2 / 3)(k_BT)^2$, to find 
\be
K =(\pi/3\hbar) k_B^2 T 
\ee
This is the upper bound on the electronic thermal conductance per
channel, or the ``quantum of thermal conductance'' (for two spin
directions). Not surprisingly (both based  on the Sommerfeld
expansion), it satisfies the Wiedemann-Franz relationship, $K = (\pi^2
/ 3)(k_B/e)^2T G$, with the conductance quantum $\frac{e^2}{\pi
  \hbar}$. 

Somewhat more surprising is the situation with the phonon thermal
conductivity. The two differences are that here heat and energy are
identical ($\mu = 0$) and that the Bose function should be used,
instead of the Fermi one. As noticed already by Bekenstein \cite{JB}
the result is nevertheless the same. The conductance quantum per
phonon mode is  $K =(\pi/6\hbar) k_B^2 T$. This was first observed
experimentally by Schwab et at \cite{Sch}. 

Finally, we turn to the upper bound on the
  thermoelectric coefficient of 1D systems. A
systematic way to get that would be to write the thermopower in terms
of the energy-dependent conductance $G(E)$ and vary the latter 
to maximize the former. However, there are Physical limitations on the
behavior of $G(E)$, which are not obvious and need more work. A
simpler approach is to appeal to the requirement of stability --
positive definiteness of the thermoelectric transport (Onsager) matrix
$\hat{{\cal L}}$, i.e., ${\cal L}_{12}^2 \le {\cal L}_{11} {\cal L}_{22}$, using the above bounds on
the diagonal elements. This gives,
\be
{\cal L}_{12} \le \sqrt{{\cal L}_{11}{\cal L}_{22}} = \frac{e}{\sqrt{3}\hbar} k_B T ,
\ee
as the upper bound for the coefficient ${\cal L}_{12}$ for spin-$\frac{1}{2}$ 1D
electron systems. As for the thermopower, it is given by ${\cal L}_{12}/(T
{\cal L}_{11})$ and it can in principle be as large as we please, for small
enough electrical conductance ${\cal L}_{11}$. As Eq.~(\ref{s-kappa}) shows,
however, when a large enough energy is transferred  by each electron,
$S$ is large. But then the output power is (usually, exponentially)
small as well (due to a small electrical conductance).

\section{Activated transport above a barrier}

\subsection{linear transport}

\begin{figure}[htb]
\includegraphics[scale=.4]{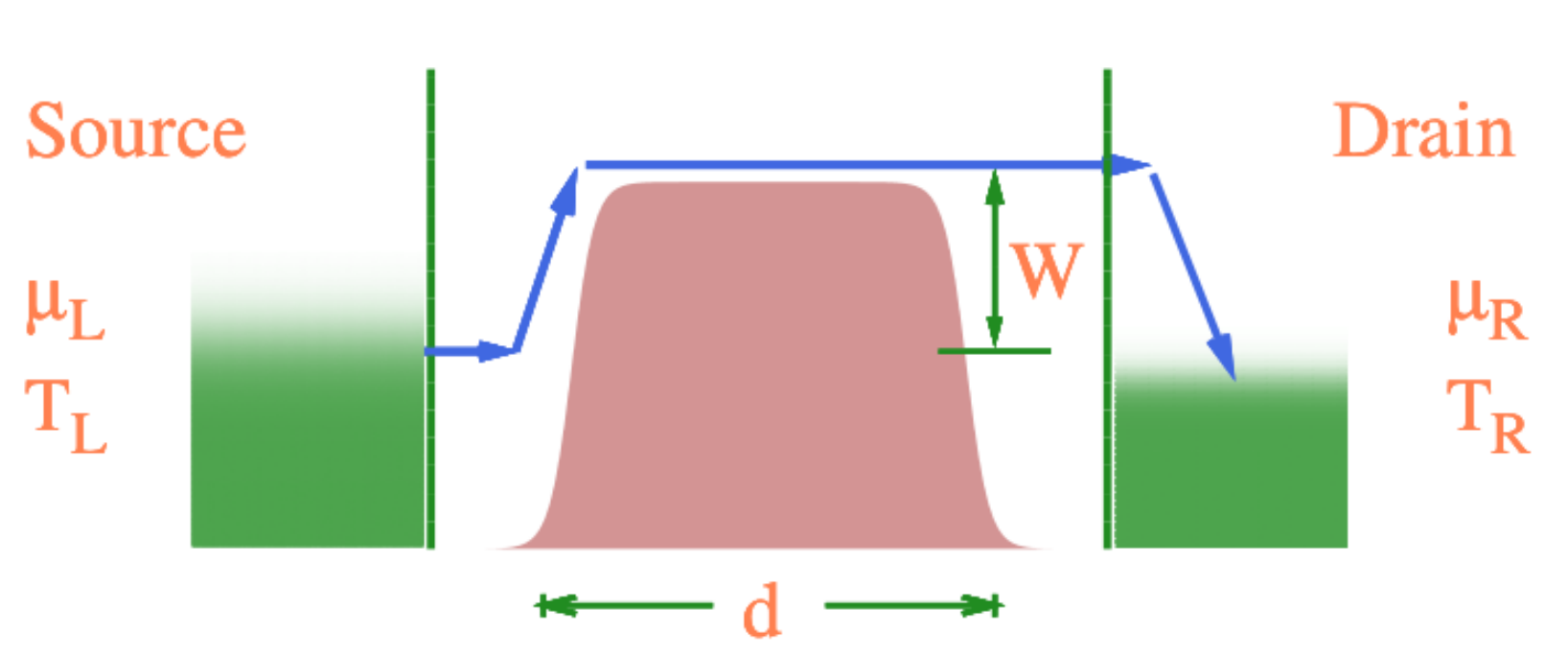}
 \centering 
   \caption{(Color online) The suggested device: a long ($d\gg$
characteristic tunneling length) and high ($W\gg k_B T$)
barrier separating two electron gases. The transferred electron gets
an energy $W\pm {\cal O}(k_BT)$ from the LHS thermal bath 
and deposits it in the RHS one.} 
\label{ba-color}
\end{figure}

Here we consider a very simple system where the electronic transport
is effectively funneled to a narrow band. We imagine an ordinary
barrier in one dimension (1D, generalized later), depicted
in Fig. \ref{ba-color}. The barrier is chosen so that its height $W$
(measured from the averaged chemical potential $\mu$) and thickness,
$d$, satisfy 
\be
W\gg k_BT; \quad d\gg \hbar /\sqrt{2mW}, 1/k_F .
\ee
The second inequality is more strongly obeyed than the first, so
that the dominant transport is via thermal activation
above the barrier and not by quantum-mechanical tunneling. We assume
that the barrier is tapered (see Fig.~\ref{ba-color}) so that the
transmission through it changes rather quickly from 0 to 1 when the
electron energy $E$ increases through $W$. This is certainly the case
in the 1D clean tunnel junction of the type discussed here, or in the
quantum point contact \cite{YI,linke}. Its validity in a high
dimensional system will be confirmed later on. When some disorder
exists, rendering the electron motion diffusive, it makes sense that
the transmission still changes from 0 to 1 when the electron energy
increases through $W$. This increase may become slower than in clean
systems, but that should not change the qualitative behavior.

A significant feature of our setup is that
the electronic heat conductance can be made to be very
small while the thermopower stays finite
\be
K_e = K_e^0 - G S^2 T^2 \ll K_e^0 ,
\ee
where $G$ is the conductance, $S$ is the thermopower, $K_e$ is the
electronic thermal conductance (defined at vanishing electrical
current) while $K_e^0$ can be termed a ``bare''
thermal conductance (defined at vanishing electrochemical potential
difference). As mentioned above, according to Ref. \cite{ms}, the largest
two-terminal figure of merit is achieved in systems with the smallest
$K_e/(GS^2T^2)$. Here this ratio is very small, and then $ZT$ is
mainly limited by the phonon heat conductivity $K_{p}$ between the
two metallic contacts [see Eq. (\ref{zt-bar}) 
below]. $K_{p}$ can be small in nanosystems\cite{nano1,nano2}. Our
system is then expected to possess a high figure of merit. Another
way to understand the situation (shown by the formulae
below) is that $S$ is the average energy transferred by an 
electron, divided by $eT$, while $K_e^0/G$ is the average of the
square of that energy, divided by $e^2$. Therefore $K_e/G$ is
proportional to the variance of that energy. The latter
obviously vanishes for a very narrow transmission band.
In this case the transmission band is the range of a few
$k_B T$ above $W$. Thus, it is not surprising that $K_e$ is of the
order of $(k_BT)^2$.

The thermoelectric linear transport problem is fully characterized, for
this two-terminal situation, by:
\be
\left( \begin{array}{c}
      I_e\\ I_{Q}^e \end{array}\right) =
  \left( \begin{array}{cccc}
      G & L_1 \\
      L_1^{'} & K_{e}^0 \\
    \end{array}\right) \left(\begin{array}{c}
      \delta\mu/e \\ \delta T/T \end{array}\right)\  .\label{2t-ba}
\ee
where $I_e$ is the charge current and $I_Q^e$ is the heat current,
$\delta T = T_L-T_R$, and $\delta\mu/e \equiv (\mu_L-\mu_R)/e \equiv V$ is the voltage between
the left and right terminals. The $2\times 2$ matrix contains
the regular conductance G, the bare electronic thermal
conductance $K_e^0$, and the (off-diagonal) thermoelectric
coefficients $L_1$ and $L_1^{'}$. That the latter two are equal is the celebrated
Onsager relation (valid for time-reversal symmetric systems). We
remind the readers that $S = L_1 /(TG)$. All currents and transport
coefficients in Eq. (\ref{2t-ba}) are given in 1D in terms of the
energy-dependent transmission of the barrier, which we take as ${\cal
  T}(E)\simeq \Theta(E-W)$. We measure all energies from the common
chemical potential $\mu$ (i.e., $\mu\equiv 0$). The currents are
\begin{align}
I_e &= \frac{2e}{h}\int_0^\infty dE {\cal T}(E) [f_L(E) - f_R(E)] ,\nn
\\
I_Q^e & = \frac{2}{h} \int_0^\infty dE E {\cal T}(E) [f_L(E) - f_R(E)] ,
\end{align}
and hence
\begin{align}
G &\simeq \frac{2e^2}{h} \int_W^\infty dE (k_BT)^{-1} f(E)[1-f(E)] \nn,\\
L_1 &\simeq \frac{2e}{h} \int_W^\infty E dE (k_BT)^{-1} f(E)[1-f(E)] \nn,\\
K_e^0 &\simeq \frac{2}{h} \int_W^\infty E^2 dE (k_BT)^{-1} f(E)[1-f(E)] ,
\end{align}
with $f(E)=1/[\exp(\frac{E}{k_BT})+1]$ being the equilibrium Fermi
distribution. It reduces for $W\gg k_BT$ to the Boltzmann distribution
and then $f(E)[1-f(E)]\propto \exp(-\frac{E}{k_BT})$. Therefore,
\begin{align}
\ave{E} = k_BT (x_w +1) , \quad 
\ave{E^2}-\ave{E}^2 = (k_BT)^2 ,
\end{align}
with $x_w \equiv W/(k_BT)$.

From the transport coefficients one readily obtains the
electronic figure of merit:
\begin{align}
ZT &= (ZT)_e \frac{K_e}{K_e+K_p} ,\label{zt-bar}\\
(ZT)_e &= \frac{\ave{E}^2}{\ave{E^2}-\ave{E}^2} = (x_w+1)^2 =
\left(\frac{W}{k_BT} +1\right)^2 .\label{zte-bar}
\end{align}
In two and three dimensions ($d = 2, 3$), the calculation
proceeds similarly. The energy is the sum of the longitudinal part
which goes over the barrier and the transverse part which should be
integrated upon. The latter has the usual density of states in $d-1$
dimensions. This is $E^{-1/2}$ for $d = 2$ and constant for $d =
3$. The overall factors do not matter for $ZT$. The final result differs
from Eq. (\ref{zte-bar}) just by numerical factors. At $d = 2$, the
result is:
\begin{align}
(ZT)_e =\frac{2}{3}\left(\frac{W}{k_BT} +\frac{3}{2}\right)^2
\end{align}
The result for $d=3$ is
\begin{align}
(ZT)_e =\frac{1}{2}\left(\frac{W}{k_BT} + 2\right)^2 .
\end{align}
The large order of magnitude of $ZT$ remains. The advantage of
the last, $d = 3$, case is twofold: it is easier to make (two thick
layers of the conducting material, separated by an appropriate
barrier) and the total current for given $W$, and hence the power, is
proportional to the cross-section of the device.

A vacuum junction has too high a barrier for most applications
\cite{Hatso}. The ballistic quantum point contact is a very effective 
realization of the model, when biased in the pinch-off regime and 
in the region where activated conduction is dominant. 
It requires, however, rather high technology and can
handle only small powers. We believe that the two more
realistic straightforward ways to effectively achieve the
requirements of the model, are:
\begin{enumerate}
\item A metal-semiconductor-metal junction, with a
properly chosen difference between work functions and
electron affinities (an example might be Au$-p$-Si$-$Au). A large area
will increase the power of the device. 
\item A superlattice \cite{shakouri} separating the two metallic
electrodes, where the Fermi level is inside the gap
between the ``valence'' and a ``conduction'' mini-bands of the
superlattice. Electron-hole symmetry should be strongly
broken either by intrinsic lack of band symmetry or
by the placement of $E_F$ away from the middle of
the gap. Obviously, highly-doped semi-
conductors can be substituted for the metallic electrodes.
\end{enumerate}

The last remark brings us to the issue of semiconducting
systems. Imagine first an intrinsic small-gap semiconductor, such as
BiTe, the current work-horse of applied thermoelectricity. The first
thought would be that since conductivity is thermally activated, with
effectively a Boltzmann distribution at temperatures much below
the gap, it is a simple realization of our model. This
would be the case only if electron-hole symmetry will
be strongly broken. If that symmetry prevails, the electron and hole
thermopowers will cancel. That symmetry breaking can be achieved by
judicious selective doping and alloying, and seems to dominate the art
of the present manipulations of BiTe and its derivatives. However,
what is proposed here is a different approach: let a larger band-gap
semiconductor, even Si, Ge, graphite, or a member of the GaAs family,
bridge two layers of metal so chosen that their Fermi level is
significantly closer to either the conduction or the valence band. In
a large temperature regime this will, as discussed in the above,
realize our activated junction model. We add, finally,
that a yet another way to realize the model is via such a
semiconductor, e.g., $n$-doped so that the Fermi level is,
say, (3 $\sim$ 10)$k_BT$ below the conduction band and much
further from the valence band. The conduction band
should then play the role of our activation barrier. Some
experimentation should show which of these schemes is
superior.

\subsection{Nonlinear activated transport above a barrier}

Here we show that a relatively high barrier, W, in the nonlinear
regime, $W  \gtrsim V \gg k_BT$, is a rather efficient thermoelectric
device. Neglecting return currents and phonon parasitic thermal
conductance, the efficiency is found to reach $W/V$, where $W$ is
the barrier the electron has to cross (setting $e\equiv 1$ in this
section). This is rather high,  especially for such a simple device. The
cooling power is $I W$, $I$ being the current. Ways to reduce the
effect of the phonon  thermal conductance are suggested.

\begin{figure}[htb]
\vspace{-.1in}
\includegraphics[height=1.8in]{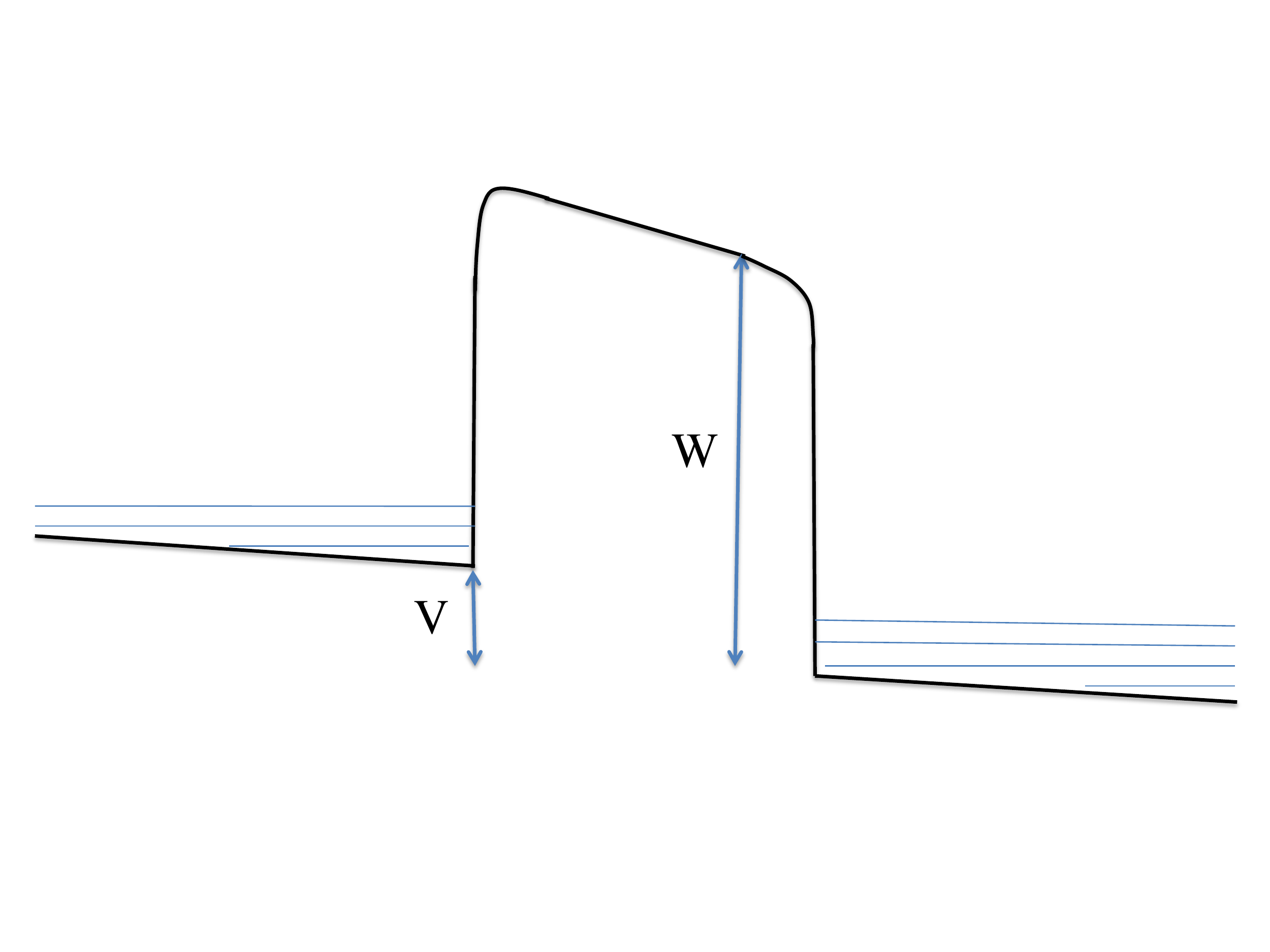}
\vspace{-.4in}
\centering 
\caption{Schematics of a barrier (W) with a large bias (V)} 
\end{figure}

We consider a wide (to neglect tunneling) solid-state or vacuum,
electronic barrier, of height $W$, bridging two conductors. It is
idealized, and taken as biased by a voltage $V$, where  $W\gtrsim V
\gg k_BT $, {\em such that all electronic conduction is from the
  left- to the right- terminal}. We have in mind $W \sim eV$.
Now, even if we keep the LHS (RHS) at temperatures $T_c$ ($T_h$), where
$T_c<T_h$, the device may transfer heat between the cool left terminal
and the warmer right one; i.e we have a (thermoelectric)
refrigerator. That heat per transported electron is given in the ideal
case roughly by $W-V\sim W \gtrsim V$  \cite{Hatso}.  This is because
an electron has to borrow an energy of about W to cross the barrier
from the LHS heat bath, it then returns it to the RHS heat bath.
Thus, the  cooling power is $IW$ and the invested power is $IV$,
where I is the current. Therefore, the efficiency and power are   
\be 
\eta = W/V ,~~~~~P = I W
\ee
\noindent where we neglected the ``parasitic'' phonon thermal
conduction back from right  to left and took the energy price per
transferred electron with the bias as $V$. 
We remark that this is not a  linear response (small $V$ and
$\delta T \equiv T_h-T_c$) calculation. It is valid for finite driving forces, as long
as $V$ is the self-consistent voltage.

By essentially reversing the process (i.e., employing temperature
differences to generate electrical energy), one may use this device
for energy harvesting as well. 

Here we give   a proof, within the linear transport approximation,
that the $W/V$ efficiency does not exceed the Carnot
one. Write the heat current as $I_{Q}^e = - K\delta T /T + L V$, where
$L=SGT$ with $S$ being the Seebeck coefficient, $K$ and $G$ 
are effective thermal and electrical conductances. The ``working
condition'' is that to refrigerate, $V > (K /L)\delta T /T
\Rightarrow$  $V/W > K/(SGWT)\delta T/ T =  K /(S^2GT^2)\delta T/ T$
(Remembering that $ST$ is the average  heat transferred by an
electron, $ S = W/T$).  Now we use the stability of the
Onsager matrix: $K/G > (ST)^2$, to get $W/V <T/\delta
T=\eta_c$ .  $\eta_c$ is the Carnot efficiency (``figure of merit'')
for a refrigerator. Thus $\eta_c$ is indeed the upper
limit on the efficiency.

The  efficiency obtained is quite high. Even pushing the voltage up
to, say, $W/2$, we get $2$ in the ideal case. If the  idealizing
factor of negligible  phonon heat conductance decreases the efficiency
by a factor of two, a value of  $1$ is still respectable for such a
simple device. It is comparable to the values for current
thermoelectric devices  having the parameter $ZT\cong1$. Of course, we
have the usual conundrum between power and efficiency. 

As far as the negative effect of the phonon thermal conductance,
besides decreasing its relative influence using a larger $V$ (decreasing
the phonons' relative contribution, see below) it is  suggested to use  a soft metal on one side of the
barrier and a hard semiconductor or metal on the other side, to reduce
the phonon thermal conductance via acoustic impedance mismatch. 
We also reemphasize that the phonon conductance is largely independent of the
applied voltage, so that {\em larger values of the latter are
  favorable also in  that respect}.

\section{The Inelastic thermoelectric transport assisted by a heat bath:
  Linear transport}

In true inelastic transport processes energy has to be
absorbed/emitted from heat baths to compensate for the energy difference
between the initial and final states. This energy can come in the form
of phonons, plasmons, charge (spin) fluctuations or other 
collective excitations. In bulk materials, those collective
excitations usually thermalize quickly with electrons. Therefore, the local
temperature of those collective excitations is nearly the same as that of the
electronic ensemble, unless the system is out of the linear-response
regime. Interaction with phonons and other collective excitations may
modify thermoelectric transport coefficients via, e.g., phonon drag effect, which
is here an unessential quantitative correction.

In mesoscopic regimes, the abundance of interfaces and disorder
may significantly reduce the thermalization processes. Substantial
temperature or electrochemical potential differences can be maintained
between adjacent nanoscale regions. Thus inelastic and nonlinear
transports become significant and prevail. Moreover, geometric
configurations that support inelastic transport are richer. It turns
out that the latter opens an important direction for improving
thermoelectric performance that has not been fully explored so
far. The fact that mesoscopic thermoelectric transport can have
significant spatial separation of electrical and thermal currents,
particularly in inelastic transport processes, enables disentanglement
of electrical and thermal conductivity and provides new opportunities
for improving the thermoelectric figure of merit.

\begin{figure}[htb]
  \includegraphics[height=1.8in]{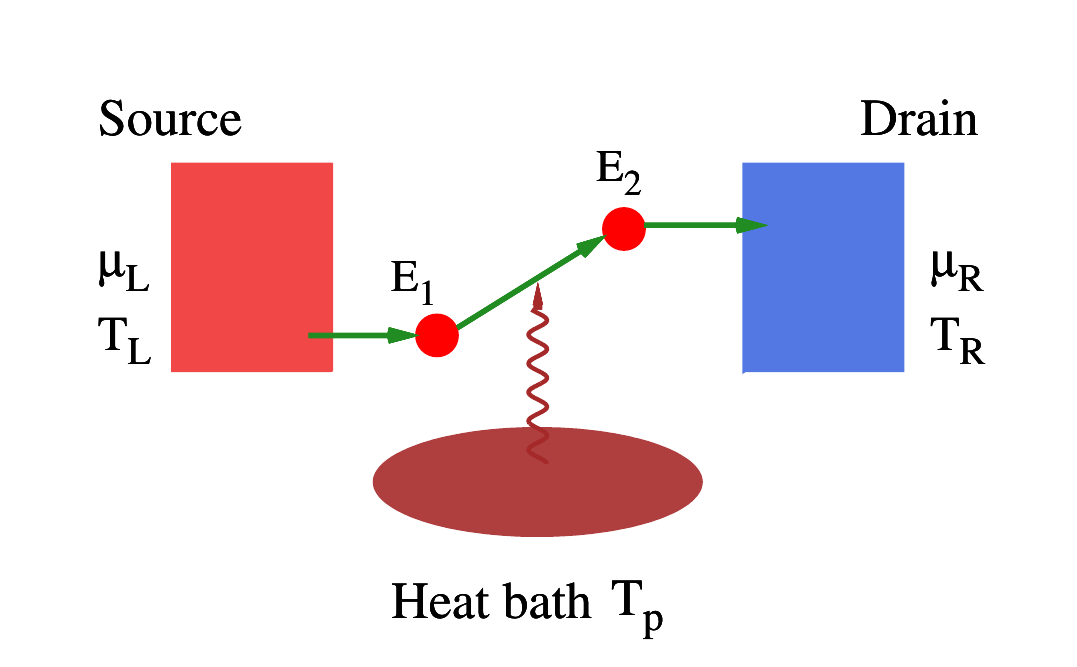}
  \centering 
  \caption{Schematic of three-terminal inelastic thermoelectric 
    mesoscopic systems. Electrons transport from source to drain have
    to absorb energy from the heat bath. There are three reservoirs:
    source, drain, and the heat bath. They can have different temperatures. }
  \label{3t-g0}
\end{figure}

The simplest nontrivial geometry configuration for inelastic transport
processes is the three-terminal set-up where energy absorbed/emitted
by the electron comes from a third heat bath differing from the source and drain
[see Fig.~\ref{3t-g0}]. Inelastic thermoelectric transport in
three-terminal geometry was first examined by Entin-Wohlman, Imry, and
Aharony\cite{joe0} for molecular junctions, and later by S\'anchez and
B\"uttiker\cite{sanchez} for capacitively coupled double quantum dots.
For the former, the additional energy is provided by a phononic heat
bath; for the latter, the additional energy is provided by a
capacitively coupled electronic reservoir. Entin-Wohlman, Imry, and
Aharony\cite{joe0} pointed out, based on thermodynamic arguments,
that the full description of thermoelectric transport in such
three-terminal system must include  two heat currents (one is the
electronic heat current from source to drain and the other is the heat
current from the third terminal), coupled with a single electrical
current. Such a property of three-terminal mesoscopic
system immediately leads to two {\em correlated} thermopowers (i.e.,
the electrical current can be induced by two temperature differences
that contribute additively in the linear-response regime). The merit
of such an effect has not been fully appreciated until recently in
Refs.~\cite{3tjap,3tjunc}. Another important property is that the
direction of heat current from the third terminal is different from 
the direction of electrical current. This means spatial separation of
electrical and thermal currents\cite{sanchez,joe0,3t,rafael-rev,jordan-rev},
which allows independent tuning of electrical and thermal conduction,
and achieve high thermoelectric figure of merit and power
factor\cite{3tjap,3tjunc}.

As in many situations, the above findings share some similarities with
some earlier works. In 1993, Edwards et al.\cite{ed} proposed to use
inelastic transport to cool an island of electrons, which can be
useful for sub-Kelvin cooling of the electron gas following conventional
cooling method\cite{ed-exp}. Nevertheless, the physics uncovered in
recent studies of mesoscopic inelastic thermoelectric effect goes far
beyond a novel method of
cooling\cite{3t,3t-nonl,3tjunc,3t-rec1,qd-engine,qw-engine,vbc,cbh}. 

The remaining of this paper will present a pedagogical review of the
essential aspects using a simple picture which is depicted in
Fig.~\ref{3t-g0}. In this 
case electrons are transported, under certain bias, from source to
drain. The dominant path for such transport is jumping elastically to a quantum
dot (or a localized energy level) with energy $E_1$ and then hopping inelastically
to another quantum dot with energy $E_2$, and finally tunneling into
the drain. Because the energies  $E_1$ and $E_2$, do not match, electrons
have to exchange energy with a heat bath. This heat bath can have a
temperature $T_p$ differing from the temperatures of the source ($T_L$)
and drain ($T_R$).

In this picture, there are three reservoirs: the source, the drain,
and the heat bath. Their states are characterized by their
temperatures and chemical potentials. The heat bath can be electronic
or phononic (or consisting of another type of bosons that carry energy).
Without loss of the essential physics, we focus on the situation that
there is no electrical charge flowing out of the heat bath. Under this
assumption we can treat the heat bath as bosonic. In this regime, the
relevant thermodynamic variable of the heat bath is just its
temperature $T_p$. The elementary excitation that provides energy to
electron transport can be phonons, charge fluctuations (i.e., charge
density waves or plasmons)\cite{3t-rec1,exp1,exp2}, spin fluctuations (i.e., spin density
waves or magnons\cite{magnon}), photons\cite{photon}, etc. This energy
is sometimes termed as energy gain in the literature\cite{sanchez}.

The energy and charge flows into the reservoirs are defined
as the time derivatives of their total energy and charge of the
reservoirs, $\dot E_i$ ($i=L,R,p$) and $e\dot N_i$
($i=L,R$) where $L$/$R$/$p$ stand for source/drain/heat bath,
respectively. $N_i$ is the total electron number of the $i$ reservoir
and $e<0$ is the charge of an electron. Total energy and charge
conservation gives $\sum_i \dot E_i=0$ and $\sum_i \dot
N_i=0$. Therefore, there are only three independent currents, which can be
organized into two heat currents and one electrical current, 
\begin{align}
I_Q^{e} \equiv \frac{1}{2}(\dot Q_R - \dot Q_L ), \quad
I_Q^{p} \equiv -\dot Q_p, \quad I_e \equiv e \dot N_R , \label{def}
\end{align}
where $\dot Q_i = \dot E_i - \mu_i \dot N_i$ for
$i=L,R$ and $\dot Q_p = \dot E_p$. 

The thermodynamic affinities conjugated to those three currents satisfy
the following relation\cite{onsager},
\be
\dot S_{tot} = I_e A_1 + I_Q^{e} A_2 + I_Q^{p} A_3 .
\ee
We found that\cite{3t-nonl} these conjugated affinities are
\begin{align}
& A_1 = \frac{\mu_L-\mu_R}{e}(\frac{1}{2T_L}+\frac{1}{2T_R}),\quad
A_{2} = \frac{1}{T_R} - \frac{1}{T_L},\nn \\
& A_{3} = \frac{1}{2T_L} +\frac{1}{2T_R} - \frac{1}{T_{p}} .
\end{align}
The phenomenological Onsager transport equation is then,
\be
I_i = \sum_j M_{ij} A_j + {\cal O} (A^2) , \label{3t-trans}
\ee
where $\hat{M}$ is the $3\times 3$ thermoelectric response matrix. In
the linear response regime,
\begin{align}
& A_1 \simeq \frac{\mu_L-\mu_R}{e T}, \quad A_{2} \simeq \frac{\delta
  T}{T^2}, \quad \delta T\equiv T_L -
  T_R, \nn\\
& A_{3} \simeq \frac{\Delta T }{T^2}, \quad \Delta T \equiv T_p - \frac{1}{2}(T_L+T_R) ,
\end{align}
where $T$ is the equilibrium temperature. And
\be
\hat{M} = T \left( \begin{array}{cccc}
    G & L_1 & L_2 \\
    L_1 & K_1 & K_{12} \\
    L_2 & K_{12} & K_2 \\
  \end{array}\right) \label{3t-dqd}
\ee
where $G$ is the conductance, $K_1$, $K_2$, and $K_{12}$ are the
diagonal and off-diagonal heat conductances. The two thermopowers are
\be
S = \frac{L_1}{TG}, \quad S_p = \frac{L_2}{TG} .
\ee
Particularly, $S_p$ enables the possibility of cooling of the heat bath by
an electrical current\cite{vbc}, or the cooling of the drain by a hot
heat bath\cite{cbh}. 

For the double quantum dots system, when inelastic
transport dominates, the transport coefficients in
Eq.~(\ref{3t-dqd}) can be written as,
\begin{align}
& L_1 = G \frac{\ov{E}}{e}, \quad L_2 = G \frac{\ome}{e},\quad K_1 = G
\frac{\ov{E}^2}{e^2},\nn\\
& K_{12} = G \frac{\ov{E}\ome}{e^2},\quad K_2 = G
\frac{\ome^2}{e^2},\nn\\
& \ov{E}\equiv \frac{E_1+E_2}{2}, \quad \ome = E_2-E_1 . 
\label{3t-dqd2}
\end{align}
The conductance is $G=\frac{e^2}{k_BT}\Gamma_{1\to 2}$ where
$\Gamma_{1\to 2}$ is the inelastic transition rate between the two
quantum dots\cite{3t,3t-nonl}. We have assumed here that the coupling
between the QD 1 and the source as well as that between the QD 2 the
drain is much stronger than the coupling between the two QDs\cite{3t,3t-nonl}.

The conventional thermoelectric effect is described by 
Eq.~(\ref{3t-trans}) for $A_3=0$, i.e., the temperature of the heat
bath is equal to the average temperature of the source and drain [more
precisely, when $T_p=2T_LT_R/(T_L+T_R)$]. Thus, $S$ is the conventional
thermopower. In contrast, $S_p$ is closely related to the inelastic
thermoelectric transport: the temperature of the heat bath comes into
play only for the inelastic thermoelectric effect. Moreover, the direction 
of heat flow out of the heat bath, $I_Q^{p}$, can be different from the
heat flow $I_Q^e$ (we shall term this as ``electronic heat current'').

In the conventional thermoelectric effect, the electrical current flows
parallel to the heat current, which leads to strong correlation
between electrical conductivity and thermal conductivity. This has
been found to be a key obstacle that limits the figure of
merit\cite{zhou}. More explicitly, the conventional thermoelectric
figure of merit here is given by
\be
ZT = \frac{GS^2T}{K_1/T-GS^2T} .
\ee
For a slab of material with area $A$ and thickness $l$, the electrical
conductivity is $\sigma = Gl/A$ and the thermal conductivity is
$\kappa = (K_1/T-GS^2T)l/A$. Hence the above equation recovers the
well-known Ioffe's figure of merit $ZT=\sigma S^2T/\kappa$.

The correlation between different transport coefficients can be
revealed by considering many parallel transport channels. For the
present model, these are many double quantum dots systems that
conduct electricity and heat between the source and the drain.
That is,
\begin{align}
& G = \int \rho(E_i) d E_i \int \rho(E_f) d E_f G(E_i,E_f), \nn\\
& S = \frac{L_1}{TG} = \frac{\ave{\ov{E}-\mu}}{eT}, \quad S_p =\frac{L_2}{TG}= \frac{\ave{\ome}}{eT}, \nn\\
& K_1 = G \frac{{\rm Var}(\ov{E}-\mu)}{e^2} + K_1^\prime, \quad K_2 = G
\frac{{\rm Var}(\ome)}{e^2} + K_2^\prime,\nn\\
& K_{12} = G \frac{\ave{(\ov{E}-\mu)\ome}}{e^2} + K_{12}^\prime , \label{3t-micro}
\end{align}
where $E_i$ is the energy of the electron when it leaves the source,
$E_f$ is its energy when it enters the drain, 
\be
\ov{E} = \frac{1}{2}(E_i+E_f), \quad \ome = E_f - E_i ,
\ee
$\rho(E)$ is the density of states, $G(E_i,E_f)$ is
the (initial and final) energy-dependent conductivity, and the average is defined as
\begin{align}
 \ave{{\cal O}} = G^{-1} \int \rho(E_i) d E_i \int \rho(E_f) d E_f
  G(E_i,E_f) {\cal O} .
\end{align}
The heat conductivities from non-electronic processes are also
included in the above. They are denoted as $K_1^\prime$, $K_2^\prime$,
and $K_{12}^\prime$, for diagonal and off-diagonal heat conduction.

We can then express the figure of merit as,
\be
ZT = \frac{\ave{\ov{E}-\mu}^2}{{\rm Var}(\ov{E}-\mu)+\Delta_1} , \quad
\Delta_1 = \frac{e^2 K_1^\prime}{G} .
\ee
The above equation, although derived from our model, also applies to
normal thermoelectric materials where $K_1^\prime$  comes from
the parasitic phonon thermal conductivity. Mahan and Sofo have argued that the
figure of merit can be significantly improved in materials with very
small electronic bandwidth $W$. In that limit, ${\rm
  Var}(\ov{E}-\mu)\sim W^2\ll (k_BT)^2$, the figure of merit is
expected to be much larger than unity if $\ave{\ov{E}-\mu}^2\sim
(k_BT)^2$. However, in the $W\to 0$ limit, the output power
vanishes\cite{low-diss}. In a recent study\cite{zhou}, it is argued that
for narrow electronic bandwidth materials, electron mobility is
suppressed by the large effective mass and strong back scattering. The
suppressed electrical conductivity $G$ then makes the factor $\Delta_1$
dominate the denominator. In such a regime the figure of merit is
suppressed, rather than promoted. Realistic optimization of the figure of
merit needs to balance the variance and the factor
$\Delta_1$\cite{zhou}. The correlation between electrical
and thermal conductivity limits further improvement of the figure of
merit\cite{zhou}.

In our system, besides the conventional thermoelectric effect, there is
an additional thermopower due to inelastic transport, $S_p$. Using
this effect, waste heat from the heat bath can be converted to useful
electrical power (reversely, electrical power can be used to cool the
heat bath without passing electricity to it). The figure of merit of
such thermoelectric energy conversion was first derived by Jiang,
Entin-Wohlman, and Imry\cite{3t,3tjunc}:
\be
\tilde{Z}T = \frac{G S_p^2 T}{K_2/T-GS_p^2T} = \frac{\ave{\ome}^2}{{\rm Var}(\ome)+\Delta_2} , \quad
\Delta_2 = \frac{e^2 K_2^\prime}{G} . \label{3t-var}
\ee
The advantages of the inelastic thermoelectric effect are manifested in
the above equation: (1) High figure of merit demands small variance of
$\ome$ (i.e., the energy of collective excitation) in the heat bath,
which can be disentangled from electronic band structures! (2) Hence
the conductivity $G$ can be uncorrelated with the variance of $\ome$
(This also opens the opportunity for high output power if $G$ is large).
(3) Sharp DOS peaks in the spectrum of collective excitations can
reduce thermal conductivity due to such excitations (i.e., reduce the
denominator in (\ref{3t-var})). This also promotes the inelastic
scattering rates and hence the electrical conductivity and the output
power. (4) Spatial separation of electrical and heat currents 
enables engineering of electrical and heat transport more
independently. For the inelastic transport to give
considerable output power and to dominate over the elastic transport,
the coupling to the bosons has to be very strong and the elastic
transport have to be suppressed by an energy barrier.

\begin{figure}[htb]
  \includegraphics[height=1.87in]{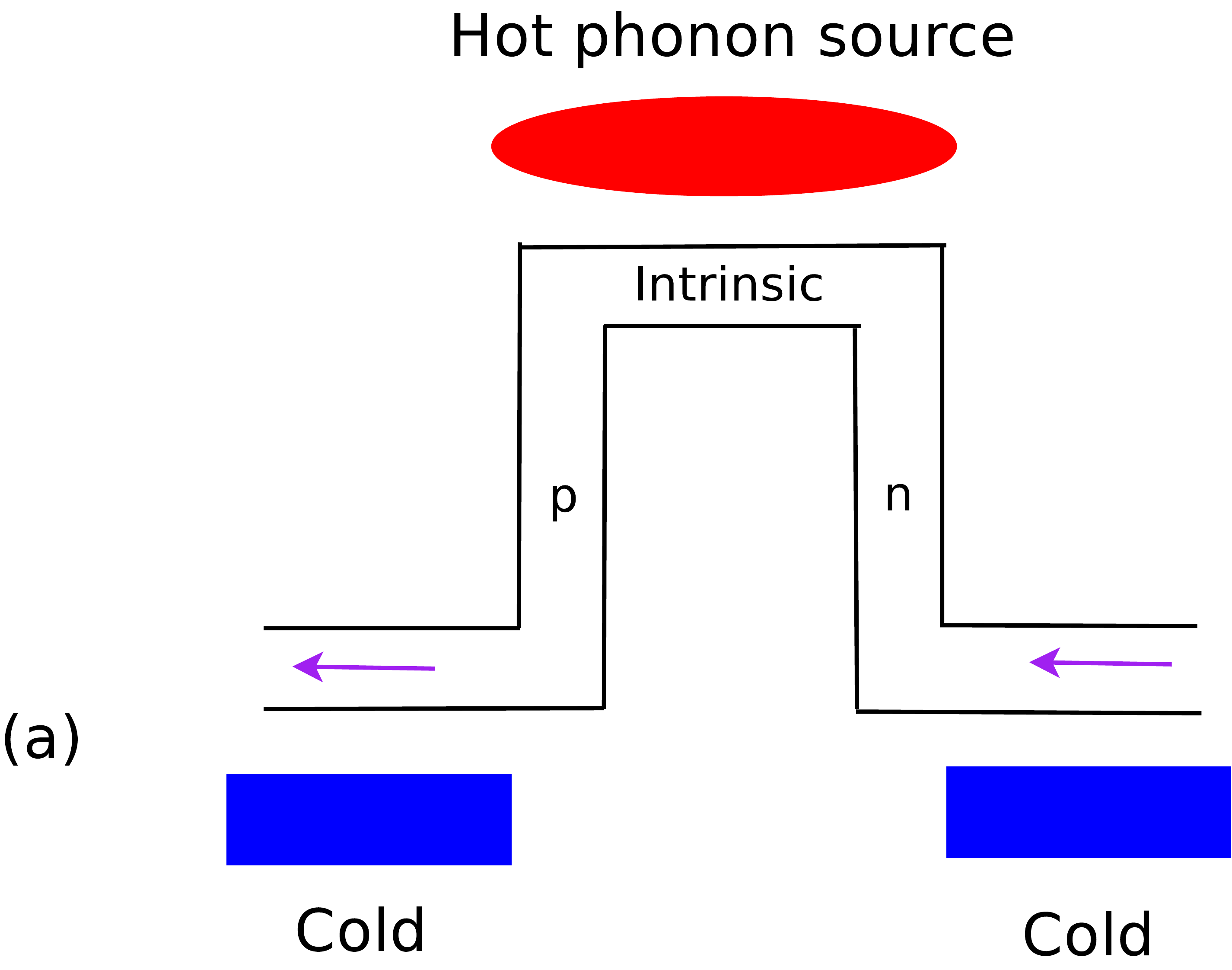}
  \vspace{.6in}
  \includegraphics[height=1.in]{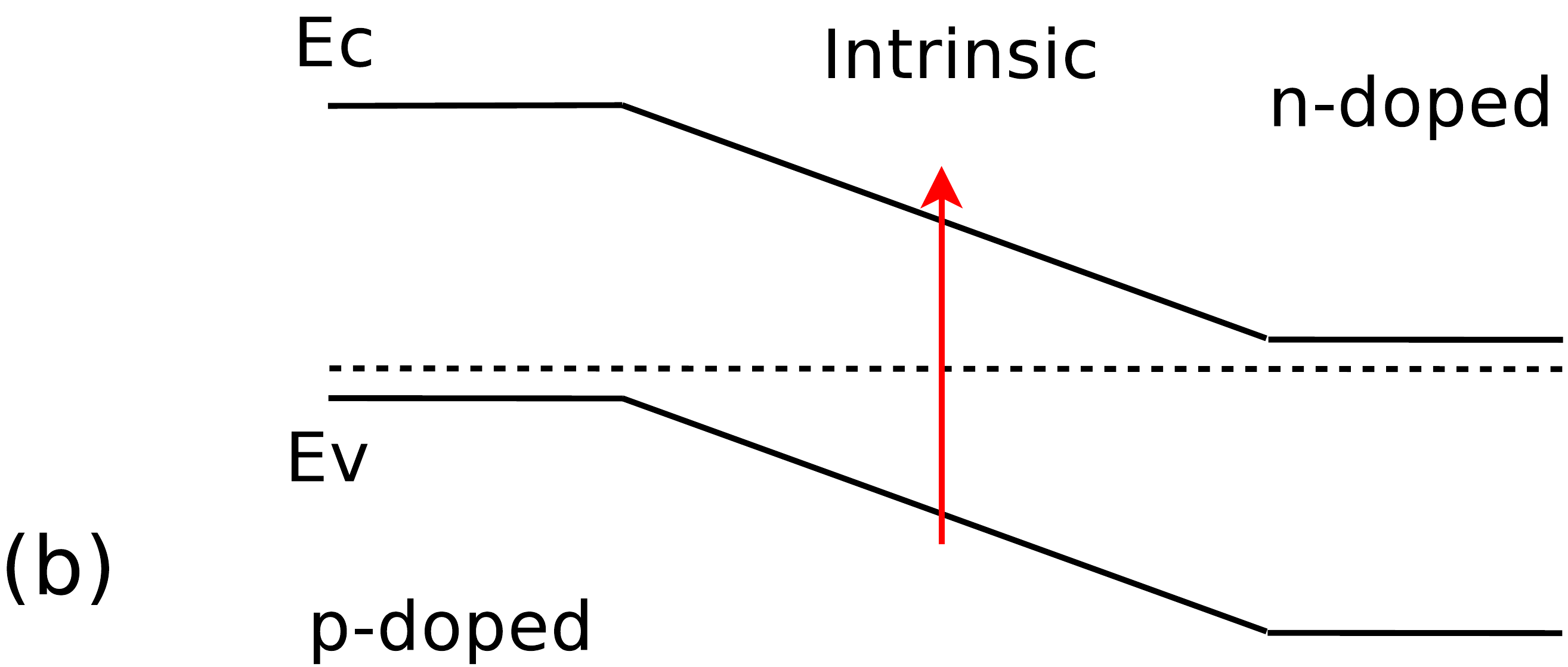}
  \centering 
  \caption{(a) Structure of $p$-$n$ junction three terminal
    thermoelectric engine. (b) Energy diagram: Absorbing a hot phonon
    generates a pair of electron and hole which drift along the
    built-in electric field of the junction to induce a current,
    similar to photo-voltaic effects. From Jiang,
      Entin-Wohlman, and Imry\cite{3tjunc}.}
  \label{pnfig}
\end{figure}

The advantages of the inelastic thermoelectric effect are explored in
Ref.~\cite{3tjunc} by considering a realistic $p$-$n$ junction
thermoelectric device. As shown in Fig.~\ref{pnfig}, a
pair of an electron and a hole is generated by absorbing a phonon from
the hot phonon reservoir. The device has similarities with a solar
cell, except that it exploits heat energy in the form of
phonons. The same paradigm can be applied for heat in the form of
other bosons, as long as the coupling between the electron and the bosons
is strong. For the case of phonons, the coupling between electrons and
optical phonons or accoustic phonons near van Hove singularities is
very strong\cite{3tjunc,haug}. Studies in Ref.~\cite{3tjunc}
indicate that, beside material
engineering, thermoelectric efficiency and power can be significantly
improved via {\em structure and geometry engineering}. Later, inelastic
thermoelectric effects based on quantum dots\cite{qd-engine,3tjap} and
quantum wells\cite{qw-engine} have been
studied. Considerably large output power and optimal efficiency have
been found with realistic quantum dot/well parameters\cite{qd-engine,3tjap,qw-engine}. In
Ref.~\cite{3tjap} the phonon thermal conductivity is included using 
experimentally measured values. It was found in Ref.~\cite{3tjap}
that, although the phonon thermal conductivity reduces both figures
of merit, $ZT$ and $\tilde{Z}T$, the inelastic thermoelectric figure
of merit $\tilde{Z}T$ is  more robust than the elastic one $ZT$.
In addition, $\tilde{Z}T$ is usually considerably larger than $ZT$
for the quantum dots array system considered in Ref.~\cite{3tjap}.
Moreover, the output power for the inelastic thermoelectric effect
is significantly greater than that of the elastic thermoelectric
effect\cite{3tjap}. The figure of merit can reach a decent value of
$\tilde{Z}T\sim 1$ using quantum dot arrays embedded in polymer thin
films\cite{3tjap} or using HgCdTe $p$-$n$
junctions\cite{3tjunc}. These results confirm 
the advantages of the inelastic thermoelectric effect.

\section{Rectification and transistor effects in the nonlinear regime
  for inelastic transport}

In this section we show how inelastic three-terminal transport
enhances nonlinear devices like diodes and transistors. Diodes and
transistors are crucial elements for electronics and conventional
information technology. There is increasing interest to invent diodes
and transistors of heat for information technology. Using heat and
electricity together to achieve information processing may have energy
advantages\cite{pre}. This idea was first explored in spin caloritronics which
tried to exploit heat to manipulate magnetic domain walls or induce
spin currents, magnon currents, that may help to reduce the energy cost
compared to using only electrical currents to achieve the same
goal\cite{bauer,pre}. A goal of pursuing the use of  heat alone for information
technology, namely ``phononics'', was also explored in the past
decade\cite{baowen}.

In Ref.~\cite{3t-nonl} it has been shown that combining thermal and
electrical degrees of freedom together, several new
nonlinear functions can be realized where
heat and electricity have synergistic effects. For example, charge
rectification can be induced by a temperature gradient, heat
rectification can be realized by voltage bias. To describe those
phenomena, we expand the currents to second order in affinities,
\be
I_i = \sum_j M_{ij} A_j + \sum_{jk} L_{ijk} A_j A_k + {\cal O} (A^3) .
\ee
The first term on the right hand side describes the linear responses,
whereas the second term gives the lowest order nonlinear responses.
The functionalities represented by various second order coefficients are
summarized in Table~I\cite{3t-nonl}. Important examples include charge
rectification induced by the temperature difference $\Delta T$, i.e.,
the difference between phonon and electronic temperature, as well as
phonon-thermoelectric transistor effect, i.e., tuning the $I_e$-$V$ or
the $I_e$-$\delta T$ curve by the phonon temperature. The $M_{ij}$ and
$L_{ijk}$ coefficients for a double quantum-dots system are calculated 
using realistic material parameters. The results in
Ref.~\cite{3t-nonl} indicate that the nonlinear effects are
considerable and can be measured using the state-of-the-art experimental
techniques. These observations indicate that synergy of
electronics and phononics  can offer new platforms and opportunities
for high performance information nanotechnologies.

\begin{table}[htb]
\caption{Functionality of second-order coefficients}
\begin{tabular}{llllllllll}\hline 
Terms ($L_{ijk}$) & \mbox{} & Diode or Transistor effect \\
\hline
$L_{111}$ &  \mbox{} & charge rectification \\
$L_{222}$, $L_{333}$ &  \mbox{} & electronic and phononic heat rectification \\
$L_{233}$, $L_{322}$ & \mbox{} & off-diagonal heat rectification \\ 
$L_{122}$, $L_{133}$ & \mbox{} & charge rectification by
$\delta T$ or $\Delta T$\\
$L_{211}$, $L_{311}$ &  \mbox{} & heat rectification by voltage\\
$L_{113}$, $L_{123}$ & \mbox{} & phonon-thermoelectric transistor \\ 
$L_{212}$, $L_{112}$ &  \mbox{} & other nonlinear thermoelectric effects\\
\hline 
\end{tabular}
\label{table1}
\end{table}

Beyond those observations, it is found that the inelastic
three-terminal transport can induce thermal transistor effect {\em in the
linear-response regime without relying on the onset of negative
differential thermal conductance}\cite{3t-nonl}. The underlying physics can be
revealed using thermodynamic arguments, without an explicit
microscopic model.

Consider the restrictions on thermal conductance in the
linear-response regime imposed by the second law of thermodynamics in
a three-terminal set-up for thermal transistor functionality.
If the thermal transport mechanisms involve only two terminals
(reservoirs) in each microscopic process, then we can write the
thermal transport equations as
\begin{subequations}
\begin{align}
& I_{L\to R} = K_{LR}(T_L - T_R)/T , \\
& I_{L\to p} = K_{Lp}(T_L - T_{p})/T , \\
& I_{R\to p} = K_{Rp}(T_R - T_{p})/T ,
\end{align}
\label{2t-trans}
\end{subequations}
where the subscripts $L$, $R$, $p$ denote the source, drain, and heat
bath, respectively, and $T$ is the equilibrium temperature. The second
law of thermodynamics requires that
\be
K_{LR} \ge 0, \quad K_{Lp}\ge 0, \quad K_{Rp} \ge 0 .
\ee

However, due to energy conservation, there are only two independent
heat currents, which we can choose as 
\be
I_Q^L = I_{L\to R} + I_{L\to p}, \quad I_Q^{p} = I_{p\to L} + I_{p\to
  R} .
\ee
Then we can cast the thermal transport equation into the form,
\begin{align}
  \left( \begin{array}{c}
      I_Q^L\\ I_{Q}^{p}\end{array}\right) =
  \left( \begin{array}{cccc}
      K_L & K_{o} \\
      K_{o} & K_{p}
    \end{array}\right) \left(\begin{array}{c}
      \frac{T_L - T_R}{T}\\ \frac{T_{p}- T_R}{T} \end{array}\right) \ .\label{3t-t}
\end{align}
where
\begin{align}
& K_L = K_{LR} +  K_{Lp} , \nn\\
&  K_{p} = K_{Lp} + K_{Rp}, \quad K_{o} = - K_{Lp} .
\end{align}
The transistor current amplification factor is defined as
\be
\alpha \equiv \left |\frac{\partial_{T_{p}} I_Q^L}{\partial_{T_{p}}
    I_Q^{p}} \right| 
\ee
In the linear-response regime one usually finds
\be
\alpha = \left|\frac{K_o}{K_{p}}\right| < 1 .
\ee
Thus no thermal transistor effect exists in the linear-response regime.

However, the above analysis missed an important type of thermal
transport mechanism that involves simultaneously {\em three} terminals 
(reservoirs) in each microscopic process. A typical example is the
inelastic transport process illustrated in Fig.~\ref{3t-g0}. For this
type of transport, the thermal transport equation can also be written
in the form of Eq.~(\ref{3t-t}). However, the second law of
thermodynamics only requires that
\be
K_L\ge 0, \quad K_{p} \ge 0, \quad K_L K_{p} \ge K_o^2 , \label{3t-res}
\ee
which does {\em not} forbid $\alpha =
\left|\frac{K_o}{K_{p}}\right|>1$. A realistic example that achieves
$\alpha>1$ in the linear-response regime can be found in
Ref.~\cite{3t-nonl}. This can also be illustrated by
considering the double quantum dot model, using
Eq.~(\ref{3t-dqd2}). Keeping in mind that
$I_Q^L=I_Q^e-\frac{1}{2}I_Q^{p}$ [using Eq.~(\ref{def}) for
$\mu_L=\mu_R$], one has
\begin{align}
\alpha = \frac{|E_1-\mu|}{|E_2-E_1|} .
\end{align}
When $|E_1-\mu|>|E_2-E_1|$ (e.g., $E_1>E_2>\mu$) then
$\alpha$ can be greater than 1 and there can be heat transistor
effect in the linear response regime.

\section{Conclusions}

Although the development of bulk semiconductor thermoelectric
materials may be limited by material parameters, the study of
mesoscopic thermoelectric effects is still far from from maturity.
It has given significant insights for improving thermoelectric
efficiency and power in nanostructured thermoelectric materials (one
of the main directions in thermoelectric research). We point out
several aspects that mesoscopic systems may help with improving
thermoelectric performances: inelastic transport, nonlinear effects,
and spatial separation of thermal and electrical currents in
thermoelectric transport. We also discuss bounds 
and quantization of thermal and thermoelectric transport coefficients
in 1D single channel conductors. Channel number and the dependence on it  are 
considered as well. We then discuss activated transport
above a barrier in both the linear and nonlinear regimes and show their
relevance for thermoelectric applications. We emphasize,
particularly, inelastic thermoelectric effects in three-terminal
geometry, which has attracted a lot of research interest recently.
The physics and merits of the three-terminal inelastic thermoelectric
effect are discussed and demonstrated for both linear and nonlinear
transport. This paper attempts to give a pedagogical review of the 
research frontiers of thermoelectric inelastic transport effects in mesoscopic
physics. Recent experimental progress\cite{exp1,exp2} is pushing
the field forward.

\vspace{0.5cm}

\section*{ACKNOWLEDGEMENTS}

We thank Hamutal Bary-Soroker, Ora Entin-Wohlman, Zvi Ovadyahu, Dan
Shahar, Moty Heiblum, Eli Zeldov and Dorri Halbertal  for useful
discussions. JHJ acknowledges support from the faculty start-up
funding of Soochow University. YI acknowledges support from
the Israeli Science Foundation (ISF) and the US-Israel Binational
Science Foundation (BSF). We thank the referee and Robert Whitney for
pointing out an error in the previous version of this paper.

{}


\begin{thebibliography}{999}

\bibitem{honig} T. C. Harman and J. M. Honig, {\sl Thermoelectric and
    thermomagnetic effects and applications} (McGraw-Hill, New-York,
  1967); H. J. Goldsmid, {\sl Introduction to Thermoelectricity} (Springer,
  Heidelberg, 2009).


\bibitem{ms} G. D. Mahan and J. O. Sofo,
  Proc. Natl. Acad. Sci. (USA) {\bf 93}, 7436 (1996). 



\bibitem{low-diss} O. Entin-Wohlman, J.-H. Jiang, and Y. Imry,
  Phys. Rev. E {\bf 89}, 012123 (2014).

\bibitem{zhou} J. Zhou, R. Yang, G. Chen, and M. S. Dresselhaus,
   Phys. Rev. Lett. {\bf 107}, 226601 (2011); C. Jeong, R. Kim, and
   M. Lundstrom, arXiv:1103.1274.


\bibitem{boundary} E.g., J.-H. Jiang, O. Entin-Wohlman, and Y. Imry,
  Phys. Rev. B {\bf 87}, 205420 (2013); R. Bosisio, C. Gorini,
  G. Fleury, and J.-L. Pichard, New J. Phys. {\bf 16}, 095005 (2014);
  R. Bosisio, C. Gorini, G. Fleury, and J.-L. Pichard,
  Phys. Rev. Applied {\bf 3}, 054002 (2015).


\bibitem{hicks} L. D. Hicks and M. S. Dresselhaus, Phys. Rev. B
  {\bf 47}, 12727 (1993); {\sl ibid.}, {\bf 47}, 16631 (1993).  


\bibitem{nano1} 
  R. Venkatasubramanian, Phys. Rev. B {\bf 61}, 3091 (2000); 
  J.-K. Yu, S. Mitrovic, D. Tham, J. Varghese, and
  J. R. Heath, Nat. Nanotechnol. {\bf 5}, 718 (2010); 
R. Venkatasubramanian,
  E. Siivola, T. Colpitts, and B. O'Quinn, Nature {\bf 413}, 597 (2001).


\bibitem{nano2} A. I. Boukai {\sl et al.}, Nature {\bf 451}, 168
  (2008); B. Poudel {\sl et al.}, Science {\bf 320}, 634 (2008);
  P. Pichanusakorn and P. Bandaru, Mater. Sci. Eng. R-Rep. {\bf 67},
  19 (2010); A. J. Minnich, M. S. Dresselhaus, Z. F. Ren, and 
  G. Chen, Energy Environ. Sci.	{\bf 2}, 466 (2009); C. J. Vineis,
  A. Shakouri, A. Majumdar, and M. G. Kanatzidis, 
  Adv. Mater. {\bf 22}, 3970 (2010); Z.-G. Chen, G. Han, L. Yang,
  L. Cheng, and J. Zou, Prog. Nat, Prog. Nat. Sci. {\bf 22}, 535
  (2012); J.-F. Li, W.-S. Liu, L.-D. Zhao, and M. Zhou, NPG Asia
  Mater. {\bf 2}, 152 (2010).


\bibitem{kanatzidis} Kanishka Biswas, Jiaqing He, Ivan D. Blum, Chun-I Wu,
  Timothy P. Hogan, David
  N. Seidman, Vinayak P. Dravid, and Mercouri G. Kanatzidis, Nature {\bf 489},
  414 (2012). 
  

\bibitem{semimetal} L. D. Hicks, T. C. Harman, M. S. Dresselhaus,
  Appl. Phys. Lett. {\bf 63}, 3230 (1993).


\bibitem{ms-review} M. S. Dresselhaus, G. Chen, M. Y. Tang, R. Yang, H.
  Lee, D. Wang, Z. Ren, J.-P. Fleurial, and P. Gogna, Adv. Mater. {\bf
  19}, 1043 (2007).



\bibitem{cutler} M. Cutler and N. F. Mott, Phys. Rev. {\bf 181},1336  (1969).

  
\bibitem{joe} U. Sivan and Y. Imry, Phys. Rev. B {\bf 33}, 551 (1986).

\bibitem{RW} R. Whitney, Phys. Rev. Lett. {\bf 112}, 130601 (2014).


\bibitem{3t} J.-H. Jiang, O. Entin-Wohlman, and Y. Imry, Phys. Rev. B
  {\bf 85}, 075412 (2012).


\bibitem{3tjap} J.-H. Jiang, J. Appl. Phys. {\bf 116}, 194303 (2014).


\bibitem{3t-nonl} J.-H. Jiang, M. Kulkarni, D. Segal, and Y. Imry,
  Phys. Rev. B {\bf 92}, 045309 (2015).



\bibitem{JB} H. G. Bremmerman, Proc. Fifth Berkeley Symp. on
  Math. Statistics and Probability, L. M, LeCam and J. Neyman, eds.
  (Univ. of Cal., Berkeley, 1966);
  J. D. Bekenstein,  Phys. Rev. Lett. {\bf 46}, 623 (1981);
  P. A. Benioff, Int. J. Theor. Phys. {\bf 21}, 177 (1982);
  J. B. Pendry, J. Phys. A: Math. and Gen. {\bf 16}, 2161 (1983).
 

\bibitem{YI} Y. Imry, {\it Physics of mesoscopic systems}. In:
   Directions in Condensed Matter, 
   Memorial volume in honor of Professor Shang-keng
   Ma. Eds. G. Grinstein and G. Mazenko (World Scientific, Singapore, 1986); 
   Y. Imry, {\it  Introduction to Mesoscopic Physics} (Oxford University Press, 
   1997 (out of print). 2nd edition,  2002).


\bibitem{Land} R. Landauer, Phil. Mag. {\bf 21}, 863 (1970).


\bibitem{exp} B. J. van Wees, H. van Houten, 
C. W. J. Beenakker, J. G. Williamson, L. P. 
Kouwenhoven, D. van der Marel, and C. T. Foxon,  
Phys. Rev. Lett., {\bf  60},  848 (1988). \\
 D. A. Wharam, T. J. Thornton, 
R. Newbury, M. Pepper, H. Ahmed, J. E. F. 
Frost, D. G. Hasko, D. C. Peacock, D. A.
 Ritchie, and G. A. C. Jones, J. Phys. C: Solid State Phys.
{\bf 21}, L209 (1988). 


\bibitem{rem1} The thermal conductance is defined under the condition
  of zero current, not zero voltage. This gives a (usually small)
  thermoelectric correction to the ``bare'' thermal conductance $K^0$
  that we calculate. We shall ignore this correction in the present section.


\bibitem{Ziman} J. M. Ziman {\it Principles of the Theory of Solids}, Chap. 7, cambridge (1969).


\bibitem{Amir} Y. Imry and A. Amir, The localization transition at finite temperatures:
electric and thermal transport, in: {\it 50 Years of Anderson Localization}, E. 
Abrahams, ed., chapter 9, pp 191-213 (World Scientific, Singapore,
2010), or arXiv:1004.0966.


\bibitem{Sch} K. Schwab, E. A. Henriksen, J. M. Worlock, and M. L. Roukes,
  Nature (London) {\bf  404}, 974  (2000). 

\bibitem{linke}  N. Nakpathomkun, H. Q. Xu, and H. Linke,
  Phys. Rev. B {\bf  82},  235428 (2010).


\bibitem{Hatso} G. D. Mahan, J. Appl. Phys. {\bf 76}, 4363 (1994);
  G. N. Hatsopoulos and E. P. Gyftopoulos, {\it Thermionic Energy
  Conversion} (MIT, Cambridge, 1973), Vol. 1. 
 

\bibitem{shakouri} D. Vashaee and A. Shakouri, Phys. Rev. Lett. {\bf
    92}, 106103 (2004).




\bibitem{joe0} O. Entin-Wohlman, Y. Imry, and A. Aharony, Phys. Rev. B
  {\bf 82}, 115314 (2010).


\bibitem{sanchez} R. S\'anchez and M. B\"uttiker, Phys. Rev. B
  {\bf 83}, 085428 (2011).



\bibitem{rafael-rev} R. S\'anchez, B. Sothmann, A. N. Jordan, and
  M. B\"uttiker, New J. Phys. {\bf 15}, 125001 (2013).

\bibitem{jordan-rev} B. Sothmann, R. S\'anchez, and A. N. Jordan,
  Nanotechnology {\bf 26}, 032001 (2015).


\bibitem{ed} H. L. Edwards , Q. Niu , and A. L. de Lozanne ,
  Appl. Phys. Lett. 63, 1815 (1993); H. L. Edwards , Q. Niu ,
  G. A. Georgakis , and A. L. de Lozanne , Phys. Rev. B 52, 5714
  (1995).

\bibitem{ed-exp} J. R. Prance, C. G. Smith, J. P. Griffiths,
  S. J. Chorley, D. Anderson, G. A. C. Jones, I. Farrer, and
  D. A. Ritchie, Phys. Rev. Lett. {\bf 102}, 146602
  (2009).


\bibitem{3tjunc} J.-H. Jiang, O. Entin-Wohlman, and Y. Imry, New
  J. Phys. {\bf 15}, 075021 (2013).


\bibitem{3t-rec1} B. Sothmann, R. S\'anchez, A. N. Jordan, and
  M. B\"uttiker, Phys. Rev. B {\bf 85}, 205301 (2012).


\bibitem{qd-engine} A. N. Jordan, B. Sothmann, R. S\'anchez, and
  M. B\"uttiker, Phys. Rev. B {\bf 87}, 075312 (2013).


\bibitem{qw-engine} B. Sothmann, R. S\'anchez, A. N. Jordan, and
  M. B\"uttiker, New J. Phys. {\bf 15} 095021 (2013).

\bibitem{vbc} L. Simine and D. Segal, Phys. Chem. Chem. Phys. {\bf
    14}, 13820 (2012); {\sl ibid.}, J. Chem. Phys. {\bf 141}, 014704 (2014);
  L. Arrachea, N. Bode, and F. von Oppen, Phys. Rev. B {\bf 90}, 125450 (2014).

\bibitem{cbh} B. Cleuren, B. Rutten, and C. Van den Broeck,
  Phys. Rev. Lett. {\bf 108}, 120603 (2012).

\bibitem{exp1} B. Roche, P. Roulleau, T. Jullien, Y. Jompol,
  I. Farrer, D.A. Ritchie, and  D.C. Glattli, Nat. Comm. {\bf 6}, 6738
  (2015).


\bibitem{exp2} H. Thierschmann et al. Nature Nanotech. {\bf 10}, 854 (2015).


\bibitem{magnon} B. Sothmann and M. B\"uttiker, Europhys. Lett. {\bf
    99} 27001 (2012).

\bibitem{photon} B. Rutten, M. Esposito, and B.
Cleuren, Phys. Rev. B {\bf 80}, 235122 (2009); T. Ruokola and
T. Ojanen, Phys. Rev. B {\bf 86}, 035454
(2012).


\bibitem{onsager} L. Onsager, Phys. Rev. {\bf 37}, 405 (1931); {\bf
    38}, 2265 (1931).

\bibitem{haug} H. Haug and S. W. Koch, {\sl Quantum Theory of the Optical
  and Electronic Properties of Semiconductors}, 4th ed. (World Scientific, Singapore, 2004).

\bibitem{pre} J. H. Jiang, Phys. Rev. E {\bf 90}, 042126 (2014).


\bibitem{bauer} G. E. W. Bauer, E. Saitoh, and B. J. van
  Wees, Nat. Mater. {\bf 11}, 391 (2012).

\bibitem{baowen} N. Li, J. Ren, L. Wang, G. Zhang, P. H\"anggi, and
  B. Li, Rev. Mod. Phys. {\bf 84}, 1045 (2012).









\end{thebibliography}
\end{document}